\documentclass[prb,reprint,twocolumn,showpacs,superscriptaddress]{revtex4-1}
\usepackage{amsmath,amssymb,bm,mathrsfs,graphicx}
\usepackage[colorlinks=true,citecolor=blue,linkcolor=blue]{hyperref}
\usepackage[usenames,dvipsnames]{color}

\begin{document}

\title{Topological Crystalline Magnets: Symmetry-Protected Topological Phases of Fermions}

\author{Haruki Watanabe}
\affiliation{Department of Applied Physics, University of Tokyo, Tokyo 113-8656, Japan.}

\author{Liang Fu}
\affiliation{Department of Physics, Massachusetts Institute of Technology, Cambridge, MA 02139, USA.}

\begin{abstract}
We introduce a novel class of interaction-enabled topological crystalline insulators in two- and three-dimensional electronic systems, which we call ``topological crystalline magnet.'' It is protected by the product of the time-reversal symmetry $\mathcal{T}$ and a mirror symmetry or a rotation symmetry $\mathcal{R}$.  A topological crystalline magnet exhibits two intriguing features: (i) it cannot be adiabatically connected to any Slater insulator and (ii) the edge state is robust against coupling electrons to the edge.  These features are protected by the anomalous symmetry transformation property $(\mathcal{R} \mathcal{T})^2=-1$ of the edge state.  An anisotropic response to the external magnetic field can be an experimental signature.
\end{abstract}

\maketitle

\section{Introduction}
Recent years have seen a great expansion of topological quantum materials beyond time-reversal-invariant topological insulators \cite{kane,zhang}, driven by the search for symmetry-protected topological (SPT) states of matter that are distinct from trivial states only in the presence of certain symmetry. This underlying symmetry can be associated with conservation of internal quantum numbers such as charge and spin \cite{ludwig, wen,senthil}, or with spatial operations such as rotation and reflection~\cite{andofu}.  Since spatial symmetry is a common property of all crystals, a wide array of topological band insulators protected by various crystal symmetries, commonly referred to as topological crystalline insulators (TCIs) \cite{fu}, has been theorized. The hallmark of a TCI is the existence of topologically protected gapless excitations on surfaces that preserve the relevant crystal symmetry. A notable class of TCIs protected by reflection symmetry was predicted and observed in the IV-VI semiconductors Sn$_{1-x}$Pb$_x$(Te,Se) \cite{hsieh,ando,story,hasan}, and the symmetry protection of the topological surface states has been demonstrated \cite{madhavan1, madhavan2, arpes}. More recently, TCIs have been generalized to band insulators with magnetic point group symmetries \cite{Sato, Fang}, nonsymmorphic symmetries~\cite{Shiozaki, VanLeeuwen,Sid,Fang,Haruki,Sato2}, and with both glide reflection and time-reversal symmetry \cite{Sato2, hourglass, Aris}. In addition, topological insulators protected by translation \cite{FuKaneMele,Stern} and magnetic translation symmetry \cite{Moore} were studied in early works. The interplay between topology and crystallography is continuing to knit together abstract mathematics and real materials.

Recently, a new type of electronic TCIs protected by reflection symmetry has been theoretically constructed \cite{Hermele}, which is enabled by electron interactions and do not exist in free fermion systems. In a broader context,  interaction-enabled topological crystalline phases were also been found in fermion superconductors \cite{Hughes} and boson insulators \cite{Oshikawa, Chen, Cirac, Xu, Kimchi, Ying, Yoshida, HermeleChen}. Such phases are now attracting wide attention, and it is of great interest to find their material realizations and experimental signatures.

In this work, we find a new class of interaction-enabled topological crystalline insulators in two and three dimensions, which are protected by time-reversal ($\mathcal{T}$) and reflection/rotation symmetry ($\mathcal{R}$), or simply the combined symmetry $\mathcal{R} \mathcal{T}$. This phase exists in systems of spin-$\frac{1}{2}$ electrons with spin-orbit interaction, and cannot be adiabatically connected to any Slater insulator in the presence of $\mathcal{R} \mathcal{T}$ symmetry. Instead, this phase admits a natural description in terms of a magnetic system of interacting spins, hence is termed ``topological crystalline magnets'' (TCMs). A distinctive feature of TCMs is the presence of gapless spin excitations on the edge parallel to the axis of reflection. These edge states exhibit strongly anisotropic response to magnetic fields in directions parallel and perpendicular to edge.

Our model for two- and three-dimensional TCMs is adiabatically connected to an array of {\it decoupled} one-dimensional symmetry-protected topological (SPT) states, on which the $\mathcal{R} \mathcal{T}$ symmetry acts as an internal anti-unitary $\mathbb{Z}_2$ symmetry. This stacking approach provides a unifying description of all previously known topological crystalline insulators \cite{Hermele}, both with \cite{IsobeFu, Furusaki} and without \cite{Fulga, Ezawa} interactions.

The one-dimensional SPT state serving as the building block of our higher dimensional TCMs apparently looks similar to, but, in fact, is remarkably different from the Affleck, Kennedy, Lieb, and Tasaki (AKLT) state~\cite{aklt1,aklt2}.  The AKLT state belongs to the Haldane phase, which is a \emph{bosonic} SPT phase protected, for example, by the dihedral ($\mathbb{Z}_2\times\mathbb{Z}_2$) symmetry or the time-reversal symmetry~\cite{Oshikawa}.  However, the Haldane phase is not a \emph{fermionic} SPT phase and is hence trivial as an electronic phase~\cite{White,Scalapino,Rosch}. Namely, when we decompose the $S=1$ spins of the AKLT model into \emph{mobile} electrons with spin-$1/2$, the ground state is adiabatically deformable into a trivial band insulator~\cite{White,Scalapino,Rosch} while keeping the dihedral and the time-reversal symmetry.  In contrast, our 1D TCM state is a robust fermionic SPT phase protected by  $\mathcal{R} \mathcal{T}$ as we shall see now.

\begin{figure}
\includegraphics[width=0.8\linewidth]{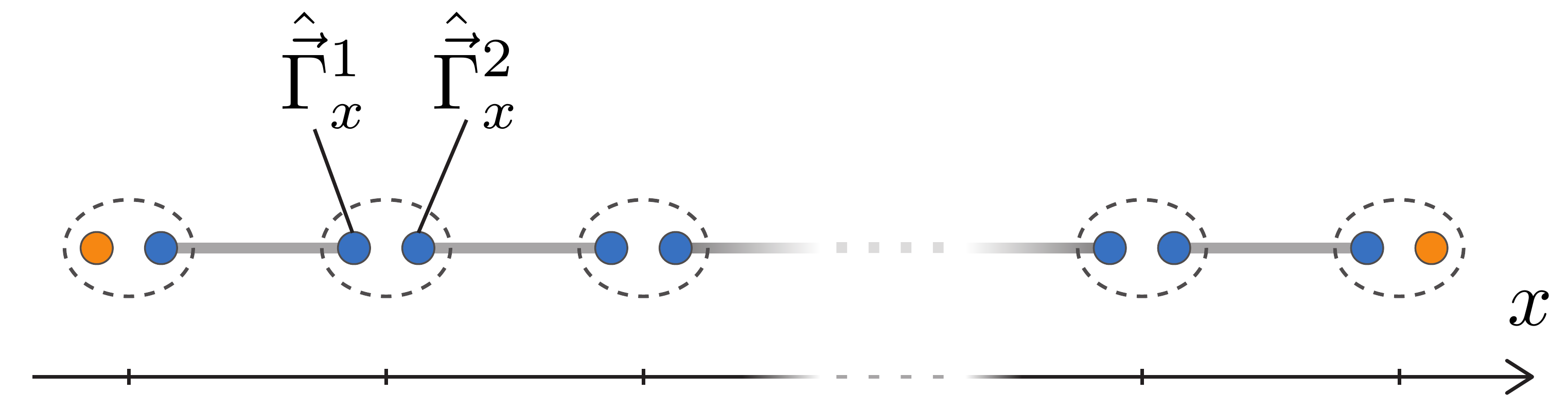}
\caption{The 1D model. Each gray bond represents a singlet pair of neighboring two $\hat{\vec{\Gamma}}_{\vec{x}}^\mu$'s and orange dots illustrate the edge degrees of freedom. A gapless edge state appears on each edge of a finite-size system. The edge degrees of freedom satisfy $(\hat{\mathcal{R}}\hat{\mathcal{T}})^2=-\hat{I}$, which is distinct from physical electrons or edge states of noninteracting topological insulators.
\label{fig}}
\end{figure}

\section{1D model}
Our 1D model (Fig.~\ref{fig}) is formed by a four-dimensional Hilbert space $\mathcal{H}_x$ on each site arising from the spin and orbital degrees of freedom of an {\it even} number of spin-$\frac{1}{2}$ electrons. The time-reversal operator $\hat{\mathcal{T}}$ thus satisfies $\hat{\mathcal{T}}^2=(-\hat{I})^{2n}=+\hat{I}$ on $\mathcal{H}_x$.  As the simplest realization of such anti-unitary symmetry we take the complex conjugation $\mathcal{T}=\mathcal{K}$.  We also assume that states in $\mathcal{H}_x$ are all even or all odd under a spatial symmetry $\mathcal{R}$, which is either the reflection about the $xz$ plane $(x,y,z) \rightarrow (x,-y,z)$ or the $\pi$-rotation about $x$-axis $(x,y,z) \rightarrow (x,-y,-z)$.  The operator $\hat{\mathcal{R}}$ is hence represented by the identity operator $\hat{\mathcal{R}}=\hat{I}$ on $\mathcal{H}_x$.  In one dimension $\mathcal{R}$ is essentially an internal symmetry, but will become a true spatial symmetry in higher dimensional cases to be studied later.

As an explicit example, $\mathcal{H}_x$ can be identified as a subset of the states of two spin-$\frac{1}{2}$ electrons occupying two orbitals. Assuming each orbital is invariant under reflection or rotation, the operator $\hat{\mathcal{R}}$ only acts on the spin part of the two-electron wavefunction. There are in total six two-electron states, consisting of spin-singlet states formed by two electrons on the same orbital, as well as spin-singlet and spin-triplet states formed by two electrons on different orbitals. We denote the electron operators associated with these two orbitals by $\hat{c}_{x,s}^\dagger$ and $\hat{d}_{x,s}^\dagger$ respectively, where $s={\uparrow},{\downarrow}$ is the spin projection along the $z$ axis. Then, out of the six two-electron states, the following four satisfy $\hat{\mathcal{R}}|n\rangle_x=(+1)|n\rangle_x$ and $\hat{\mathcal{T}}|n\rangle_x=(+1)|n\rangle_x$ ($n=1,2,3,4$) and span the desired Hilbert space $\mathcal{H}_x$:
\begin{eqnarray}
|1\rangle_x&\equiv&\hat{c}_{x\uparrow}^\dagger \hat{c}_{x\downarrow}^\dagger |0\rangle,\\
|2\rangle_x&\equiv&\hat{d}_{x\uparrow}^\dagger \hat{d}_{x\downarrow}^\dagger |0\rangle,\\
|3\rangle_x&\equiv&\frac{1}{\sqrt{2}}(\hat{c}_{x\uparrow}^\dagger \hat{d}_{x\downarrow}^\dagger- \hat{c}_{x\downarrow}^\dagger \hat{d}_{x\uparrow}^\dagger) |0\rangle,\\
|4\rangle_x&\equiv&\frac{1}{\sqrt{2}}(\hat{c}_{x\uparrow}^\dagger \hat{d}_{x\uparrow}^\dagger+ \hat{c}_{x\downarrow}^\dagger \hat{d}_{x\downarrow}^\dagger) |0\rangle.
\end{eqnarray}
The remaining two states can also be included in the following discussion, but as long as their energy level is set much higher than these four states, they will not affect the topological property of our ground state.

The 1D Hamiltonian for a finite chain $1\leq x\leq L$ reads
\begin{equation}
\hat{H}_{\text{1D}}=J\sum_{x=1}^{L-1}\hat{\vec{\Gamma}}^1_x \cdot \hat{\vec{\Gamma}}^2_{x+1},\label{1D}
\end{equation}
where both $\hat{\vec{\Gamma}}^1$ and $\hat{\vec{\Gamma}}^2$ are a set of three Hermitian operators that generate the $SU(2)$ algebra and mutually commute, i.e.,
\begin{eqnarray}
[\hat{\Gamma}^{\mu a},\hat{\Gamma}^{\mu b}]&=&i \epsilon^{abc}\hat{\Gamma}^{\mu c}, \;\; [\hat{\Gamma}^{1 a}, \hat{\Gamma}^{2 b}] =0
\end{eqnarray}
with $a,b=x,y,z$ and $\mu=1,2$.
The components of these $\Gamma$ operators are explicitly given by the following $4\times4$ matrices in the basis of $|n\rangle$
\begin{eqnarray}
\vec{\Gamma}^1 &\equiv& \frac{1}{2}(-\sigma^z\otimes\sigma^y,-\sigma^y\otimes\sigma^0,-\sigma^x\otimes\sigma^y),\\
\vec{\Gamma}^2&\equiv& \frac{1}{2}(-\sigma^0\otimes\sigma^y,\sigma^y\otimes\sigma^z,-\sigma^y\otimes\sigma^x).
\end{eqnarray}
Note that $\vec{\Gamma}^{1,2}$ are pure imaginary and are hence odd under time-reversal symmetry $\mathcal{T}$. The Hamiltonian \eqref{1D} consists of bilinears of $\Gamma$'s and is therefore time-reversal invariant. It is also invariant under $\mathcal{R}$ since $\mathcal{R}$ does not transform $\Gamma$ at all.

To analyze the topological nature of the ground state of $\hat{H}_{\text{1D}}$, it is more convenient to switch the basis of $\mathcal{H}_x$ from $\{|n\rangle_x \}_{n=1,2,3,4}$ to $\{|s\rangle_x^1\otimes|s'\rangle_x^2\}_{s,s'=\pm}$ by the local linear transformation $|s\rangle_x^1\otimes|s'\rangle_x^2=\sum_{n}|n\rangle_x U_{n,ss'}$:
\begin{eqnarray}
|+\rangle_x^1\otimes|+\rangle_x^2&\equiv&\frac{1}{\sqrt{2}}(|1\rangle_x-i|4\rangle_x),\label{trans1}\\
|+\rangle_x^1\otimes|-\rangle_x^2&\equiv&\frac{1}{\sqrt{2}}(|3\rangle_x-i|2\rangle_x),\\
|-\rangle_x^1\otimes|+\rangle_x^2&\equiv&-\frac{1}{\sqrt{2}}(|3\rangle_x+i|2\rangle_x),\\
|-\rangle_x^1\otimes|-\rangle_x^2&\equiv&\frac{1}{\sqrt{2}}(|1\rangle_x+i|4\rangle_x).\label{trans4}
\end{eqnarray}
In this new basis, $\hat{\vec{\Gamma}}_x^\mu$ is nothing but the spin operator acting on $\{|s\rangle_x^\mu\}_{s=\pm}$,
\begin{equation}
U^\dagger\vec{\Gamma}^1U=\frac{1}{2}\vec{\sigma}\otimes\sigma^0,\quad U^\dagger\vec{\Gamma}^2U=\frac{1}{2}\sigma^0\otimes\vec{\sigma}.\label{spin}
\end{equation}
For example, the usual spin algebras such as $\hat{\Gamma}_x^{\mu z}|\pm\rangle_x^\mu=\pm\frac{1}{2}|\pm\rangle_x^{\mu}$ and $(\hat{\Gamma}_x^{\mu x}\pm i\hat{\Gamma}_x^{\mu y})|\mp\rangle_x^{\mu}=|\pm\rangle_x^\mu$ hold.
Therefore, $\hat{H}_{\text{1D}}$ in Eq.~\eqref{1D} is just an antiferromagnetic spin chain whose exchange coupling is nonzero in every other bond.  The ground state is the valence-bond solid (VBS) state:
\begin{eqnarray}
|\Psi(s,s')\rangle&\equiv&|s\rangle_1^1\otimes\left(\Pi_{x=1}^{L-1} \otimes|\phi_0\rangle_{x,x+1}\right)\otimes|s'\rangle_L^2,\label{VBS}\\
|\phi_0\rangle_{x,x+1}&\equiv&\frac{1}{\sqrt{2}}(|+\rangle_x^2|-\rangle_{x+1}^1-|-\rangle_x^2|+\rangle_{x+1}^1).
\end{eqnarray}
In a finite-size system, the ground state is four-fold degenerate due to the edge dofs $|s\rangle_1^1$ and $|s'\rangle_L^2$ ($s,s'=\pm$).

The nontrivial topology of the model is encoded in the symmetry property of the edge states.
Although the auxiliary field $|s\rangle_x^\mu$ apparently behaves like an electronic spin, its transformation under $\hat{\mathcal{R}}\hat{\mathcal{T}}$ is in fact quite distinct from the physical spin.  In the $\{|s\rangle_x^1\otimes|s'\rangle_x^2\}$ basis, $\hat{\mathcal{T}}$ and $\hat{\mathcal{R}}$ are represented by $U^\dagger \mathcal{T} U=U^\dagger U^*\mathcal{K}=(i\sigma^y)\otimes(i\sigma^y)\mathcal{K}$ and $U^\dagger \mathcal{R} U=U^\dagger IU=\sigma^0\otimes\sigma^0$, respectively.  Namely, $|s\rangle_x^\mu$ transforms under $\mathcal{T}$ in the same way as the physical spins, while it does not change under $\mathcal{R}$ ($\hat{\mathcal{R}}|\pm\rangle_x^\mu=|\pm\rangle_x^\mu$) unlike electrons.  This peculiar transformation property of the auxiliary field $|s\rangle_x^\mu$ can be summarized as
\begin{equation}
\hat{\mathcal{R}}=+\hat{I},\quad \hat{\mathcal{T}}^2=-\hat{I}, \quad (\hat{\mathcal{R}}\hat{\mathcal{T}})^2=-\hat{I}
\label{RT2}
\end{equation}
on the two dimensional Hilbert space spanned by $\{|s\rangle_x^\mu\}_{s=\pm}$.  Equation~\eqref{RT2} must be compared to $\hat{\mathcal{T}}^2=\hat{\mathcal{R}}^2=-\hat{I}$ and hence $(\hat{\mathcal{R}}\hat{\mathcal{T}})^2=+\hat{I}$ of a physical spin-$\frac{1}{2}$ electron.  One may think one can redefine $\hat{\mathcal{R}}'\equiv i \hat{\mathcal{R}}$ to get $(\hat{\mathcal{R}}')^2=+\hat{I}$, but even after that $(\hat{\mathcal{R}}'\hat{\mathcal{T}})^2$ remains unchanged since $\hat{\mathcal{T}}$ is anti-unitary.

Although the Hamiltonian $\hat{H}_{\text{1D}}$ is invariant under $\hat{\mathcal{T}}$ and $\hat{\mathcal{R}}$ separately, we can add arbitrary symmetry-breaking perturbations keeping only the combined symmetry $\hat{\mathcal{R}}\hat{\mathcal{T}}$ and the bulk gap.  Since $\hat{\mathcal{R}}\hat{\mathcal{T}}$ is an anti-unitary symmetry that squares into $-1$, it protects the Kramers degeneracy on each edge.

The fact that the value of $(\hat{\mathcal{R}}\hat{\mathcal{T}})^2$ of our edge state is different from that of physical electrons has two important implications. (i) The edge state of any (noninteracting) topological insulator satisfies $(\hat{\mathcal{R}}\hat{\mathcal{T}})^2=+\hat{I}$.  Therefore, the VBS state in Eq.~\eqref{VBS} cannot be adiabatically connected to electronic topological insulators.  In other words, the VBS state is an interaction-enabled topological phase protected by $\hat{\mathcal{R}}\hat{\mathcal{T}}$.  (ii) The edge state of the VBS state is robust against the perturbation of attaching physical spin-$\frac{1}{2}$ electrons to the edge.  In the case of the standard AKLT model, for example, the edge spin-$\frac{1}{2}$ can be gapped by attaching an electron, since both of them fall into the same class of projective representations $\hat{\mathcal{R}}^2=\hat{\mathcal{T}}^2=-\hat{I}$.  On the other hand, the edge state of our model cannot be gapped this way, since even after attaching an electron, the anti-unitary symmetry $\hat{\mathcal{R}}\hat{\mathcal{T}}$ remains $(\hat{\mathcal{R}}\hat{\mathcal{T}})^2=-\hat{I}$.

To summarize, we have presented a simple 1D model of interacting electrons that realizes an interaction-enabled topological phase protected by the combined symmetry $\hat{\mathcal{R}}\hat{\mathcal{T}}$. The edge degrees of freedom satisfy $(\hat{\mathcal{R}}\hat{\mathcal{T}})^2=-\hat{I}$ and are stable against attaching additional electrons to the edge.  

\begin{figure}
\includegraphics[width=0.9\linewidth]{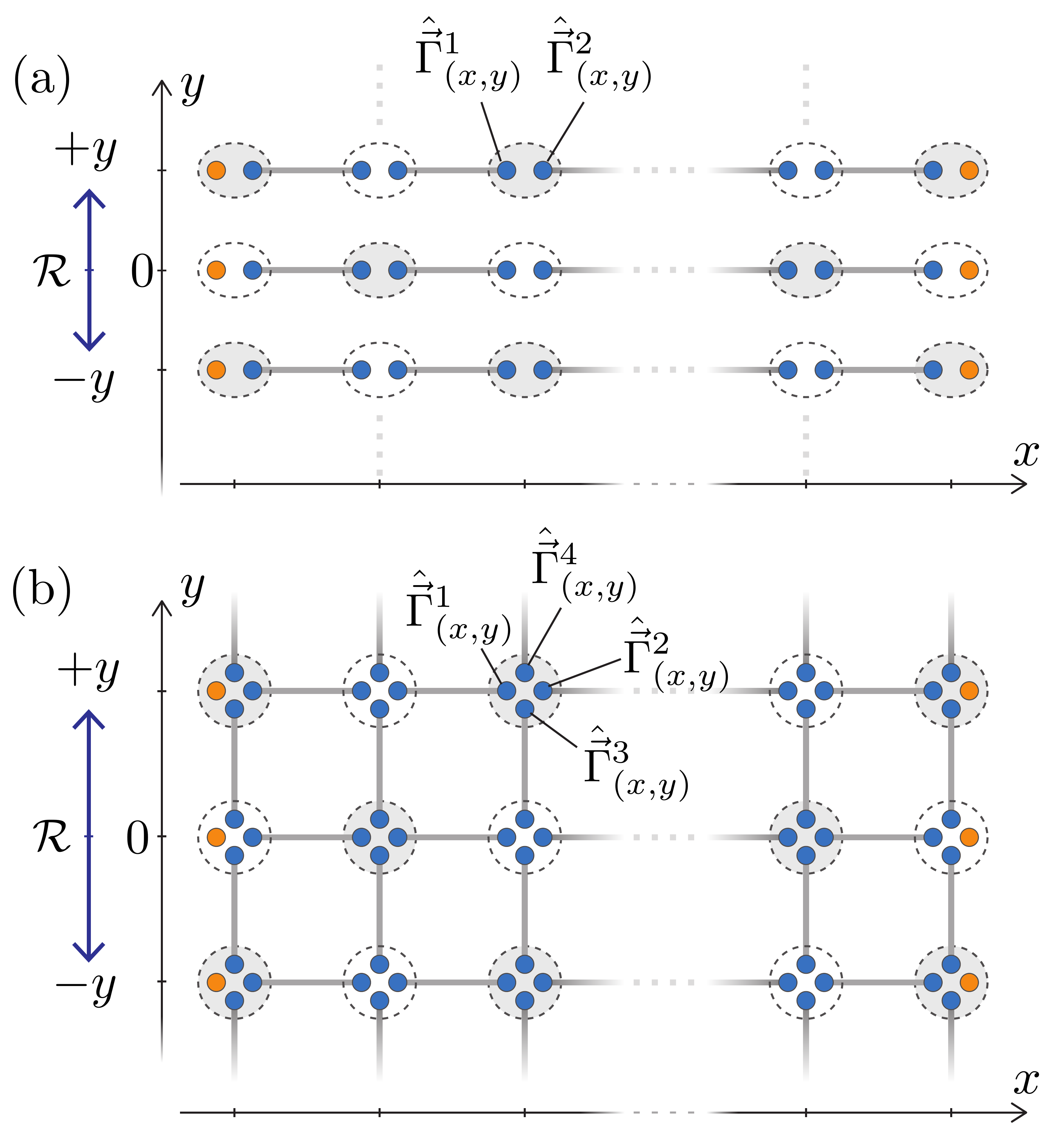}
\caption{Two 2D models: the stacked 1D chains (a) and a more intrinsically 2D model (b).  The reflection/rotation symmetry must be site-centered, not bond-centered.  We can add weak perturbation to realize A-B sublattice structure (gray shadow) to break the bond-centered mirror.
\label{fig2}}
\end{figure}

\section{2D models}
Now we move onto 2D TCM models. This time the reflection/rotation symmetry is truly a spatial symmetry and the 2D TCM phases are hence protected purely by non-local symmetries.

We will discuss two models. The first one is stacked 1D chains shown in Fig.~\ref{fig2} (a). The Hamiltonian is
\begin{equation}
\hat{H}_{\text{2D}}=J\sum_{x=1}^{L_x-1}\sum_{y=-\infty}^{+\infty}\hat{\vec{\Gamma}}_{(x,y)}^2\cdot\hat{\vec{\Gamma}}_{(x+1,y)}^1,\label{2D1}
\end{equation}
where $\vec{\Gamma}^1\equiv\frac{1}{2}\vec{\sigma}\otimes\sigma^0$ and $\vec{\Gamma}^2\equiv\frac{1}{2}\sigma^0\otimes\vec{\sigma}$ in the basis of $\{|s_1\rangle_{(x,y)}^1\otimes|s_2\rangle_{(x,y)}^2\}_{s_1,s_2=\pm}$.
The second one is a square-lattice model depicted in Fig.~\ref{fig2} (b).
\begin{eqnarray}
\hat{H}_{\text{2D}}'=&&J\sum_{x=1}^{L_x-1}\sum_{y=-\infty}^{+\infty}\hat{\vec{\Gamma}}_{(x,y)}^2\cdot\hat{\vec{\Gamma}}_{(x+1,y)}^1\notag\\
&&+J\sum_{x=1}^{L_x}\sum_{y=-\infty}^{+\infty}\hat{\vec{\Gamma}}_{(x,y)}^4\cdot\hat{\vec{\Gamma}}_{(x,y+1)}^3,\label{2D2}
\end{eqnarray}
where $\vec{\Gamma}^1\equiv\frac{1}{2}\vec{\sigma}\otimes\sigma^0\otimes\sigma^0\otimes\sigma^0$, $\vec{\Gamma}^2\equiv\frac{1}{2}\sigma^0\otimes\vec{\sigma}\otimes\sigma^0\otimes\sigma^0$, $\vec{\Gamma}^3\equiv\frac{1}{2}\sigma^0\otimes\sigma^0\otimes\vec{\sigma}\otimes\sigma^0$, and $\vec{\Gamma}^4\equiv\frac{1}{2}\sigma^0\otimes\sigma^0\otimes\sigma^0\otimes\vec{\sigma}$ in the basis of $\{|s_1\rangle_{(x,y)}^1\otimes|s_2\rangle_{(x,y)}^2\otimes|s_3\rangle_{(x,y)}^3\otimes|s_4\rangle_{(x,y)}^4\}_{s_1,s_2,s_3,s_4=\pm}$.

For both models, each auxiliary field $|s\rangle_{(x,y)}^\mu$ ($s=\pm$) transforms as
\begin{equation}
\hat{\mathcal{R}}|s\rangle_{(x,y)}^\mu=|s\rangle_{(x,-y)}^\mu,\quad \hat{\mathcal{T}}|s\rangle_{(x,y)}^\mu=s\,|-s\rangle_{(x,y)}^\mu
\label{RT2D}
\end{equation}
so that $\hat{\vec{\Gamma}}_{(x,y)}^\mu$ satisfies
\begin{equation}
\hat{\mathcal{R}}\hat{\vec{\Gamma}}_{(x,y)}^\mu\hat{\mathcal{R}}^{-1}=\hat{\vec{\Gamma}}_{(x,-y)}^\mu,\,\,\hat{\mathcal{T}}\hat{\vec{\Gamma}}_{(x,y)}^\mu\hat{\mathcal{T}}^{-1}=-\hat{\vec{\Gamma}}_{(x,y)}^\mu.
\label{Gamma2D}
\end{equation}
The first transformation in Eq.~\eqref{RT2D} is again distinct from that of spin-$\frac{1}{2}$ electrons.  As a consequence, $|s\rangle^{\mu}_{(x,y)}$ satisfies $(\hat{\mathcal{R}}\hat{\mathcal{T}})^2=-\hat{I}$ unlike electrons as before. Although both $\hat{H}_{\text{2D}}$ and $\hat{H}_{\text{2D}}'$ themselves are invariant under $\hat{\mathcal{R}}$ and $\hat{\mathcal{T}}$ separately, arbitrary perturbations can be added to these Hamiltonians as long as the combined symmetry $\hat{\mathcal{R}}\hat{\mathcal{T}}$ is respected and the bulk gap is not closed.

Note that the reflection/rotation symmetry $\mathcal{R}$ here needs to be site-centered [$\mathcal{R}:(x,y,z)\mapsto(x,-y,\pm z)$] and cannot be bond-centered [$\tilde{\mathcal{R}}:(x,y,z)\mapsto(x,1-y,\pm z)$]. The bond-centered one does not protect gapless edge states as we discuss below.  To break the bond-centered symmetry without affecting the site-centered one,  one can introduce A-B sublattice structure [gray shadows in Fig.~\ref{fig}(b)] by modifying the spin Hamiltonian by weak perturbation.

\begin{figure}
\includegraphics[width=\linewidth]{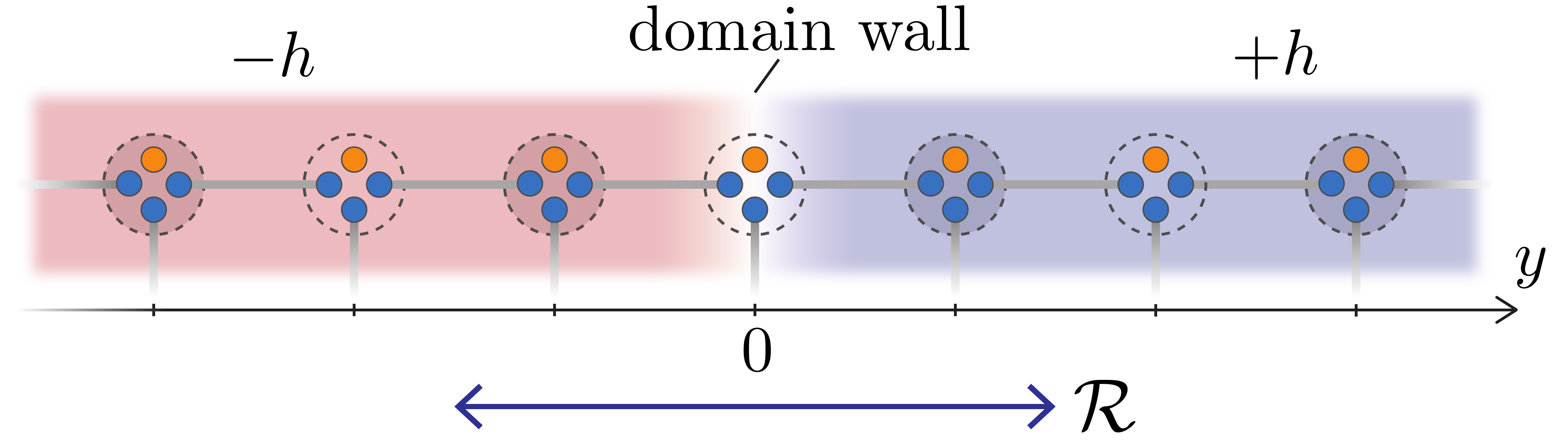}
\caption{The 1D edge state along $x=1$ of the 2D models in Fig.~\ref{fig2}.  The color represents the nonuniform perturbation $\hat{H}'=\sum_{y} \vec{H}(y)\cdot\hat{\vec{\Gamma}}_{(x=1,y)}^1$ with $\vec{H}(y)=h\,\text{tanh}(y) \hat{z}$, for example, which respects the $\hat{\mathcal{R}}\hat{\mathcal{T}}$ symmetry.  The edge state in the region $y\gg0$ and $y\ll0$ opens a gap proportional to $h$, while there will be a residual gapless edge state protected by $(\hat{\mathcal{R}}\hat{\mathcal{T}})^2=-\hat{I}$ on the domain wall.
\label{fig3}}
\end{figure}

The ground state of these 2D Hamiltonians is the VBS state illustrated in Fig.~\ref{fig2}, analogous to Eq.~\eqref{VBS}.  There is a 1D edge state formed by $\{|s\rangle_{(x=1,y)}^1\}_{y\in[-L_y,L_y]}$ along the line $x=1$, and another 1D edge state formed by $\{|s\rangle_{(x=L_x,y)}^2\}_{y\in[-L_y,L_y]}$ along $x=L_x$.

To see the gaplessness of the edge states, we add a $\hat{\mathcal{R}}\hat{\mathcal{T}}$-symmetric perturbation $\hat{H}'=\sum_{y} \vec{H}(y)\cdot\hat{\vec{\Gamma}}_{(x=1,y)}^1$ along the line $x=1$ as shown in Fig.~\ref{fig3}, where $\vec{H}(y)$ is an odd function of $y$ that approaches to a constant $\vec{H}(y)\rightarrow\vec{h}$ for $y\gg1$.  Note that $\vec{H}(y)$ must flip sign at $y=0$ to be consistent with the $\hat{\mathcal{R}}\hat{\mathcal{T}}$ symmetry, forming a domain wall around $y=0$.  All $\hat{\vec{\Gamma}}$'s along the edge away from the domain wall open a gap proportional to $h$.  However, the edge state at the domain wall $y=0$ must remain gapless. This is protected, again, by the anti-unitary symmetry $\hat{\mathcal{R}}\hat{\mathcal{T}}$ with $(\hat{\mathcal{R}}\hat{\mathcal{T}})^2=-\hat{I}$.   This unavoidable gaplessness of the edge state signals the topological nature of our 2D models.  Essentially, $\vec{\Gamma}_{\vec{x}}^\mu$'s on the $y=z=0$ line play the role of the 1D spin chain discussed above.  In contrast, when $\mathcal{R}$ is bond-centered, there will be an even number of $|s\rangle$'s at the domain wall and $(\hat{\mathcal{R}}\hat{\mathcal{T}})^2=(-\hat{I})^{2n}=+\hat{I}$ and the edge may be completely gapped.

\section{Anisotropic response to a magnetic field}
An experimental signature of TCMs is the anisotropic response of the edge state to the external magnetic field $\vec{B}$.

We start with the case where $\mathcal{R}$ is the reflection $\mathcal{M}_y$ about the $xz$ plane.  Recall that the $(B_x, B_y, B_z)\rightarrow (-B_x, B_y, -B_z)$ under $\mathcal{M}_y$, while $\hat{\vec{\Gamma}}$ does not react to $\mathcal{M}_y$.  Both $\vec{B}$ and $\hat{\vec{\Gamma}}$ flip sign under $\mathcal{T}$. The familiar form of the coupling to the external field $\vec{B}\cdot\hat{\vec{\Gamma}}_{\vec{x}}^{\mu}$ is thus not allowed by symmetry $\mathcal{M}_y\mathcal{T}$.  Instead, arbitrary linear coupling to $B_y$, i.e., $B_yc_{\mu a}\hat{\Gamma}_{\vec{x}}^{\mu a}$, is allowed. When $B_y$ is set to a constant value, this term breaks the $\mathcal{M}_y\mathcal{T}$ symmetry and the edge states will be gapped and the gap should be proportional to $|B_y|$.  On the other hand, $B_{x}$ and $B_{z}$ do not couple linearly to $\hat{\Gamma}_{\vec{x}}^{\mu}$.  We therefore expect anisotropic response of the edge state towards the external magnetic field.

When $\mathcal{R}$ is the $\pi$-rotation $\mathcal{R}_{\pi,x}$ around $x$ axis, the magnetic field $(B_x, B_y, B_z)$ changes to $(B_x, -B_y, -B_z)$ under $\mathcal{R}_{\pi,x}$.  Thus, arbitrary linear coupling between $B_x$ and $\Gamma_{\vec{x}}^{\mu a}$ is allowed.  Thus a constant $B_x$ can induce a gap to the edge, while $B_y$ and $B_z$ cannot. We thus expect similar anisotropic response in this case too.

\begin{figure}[b]
\includegraphics[width=0.6\linewidth]{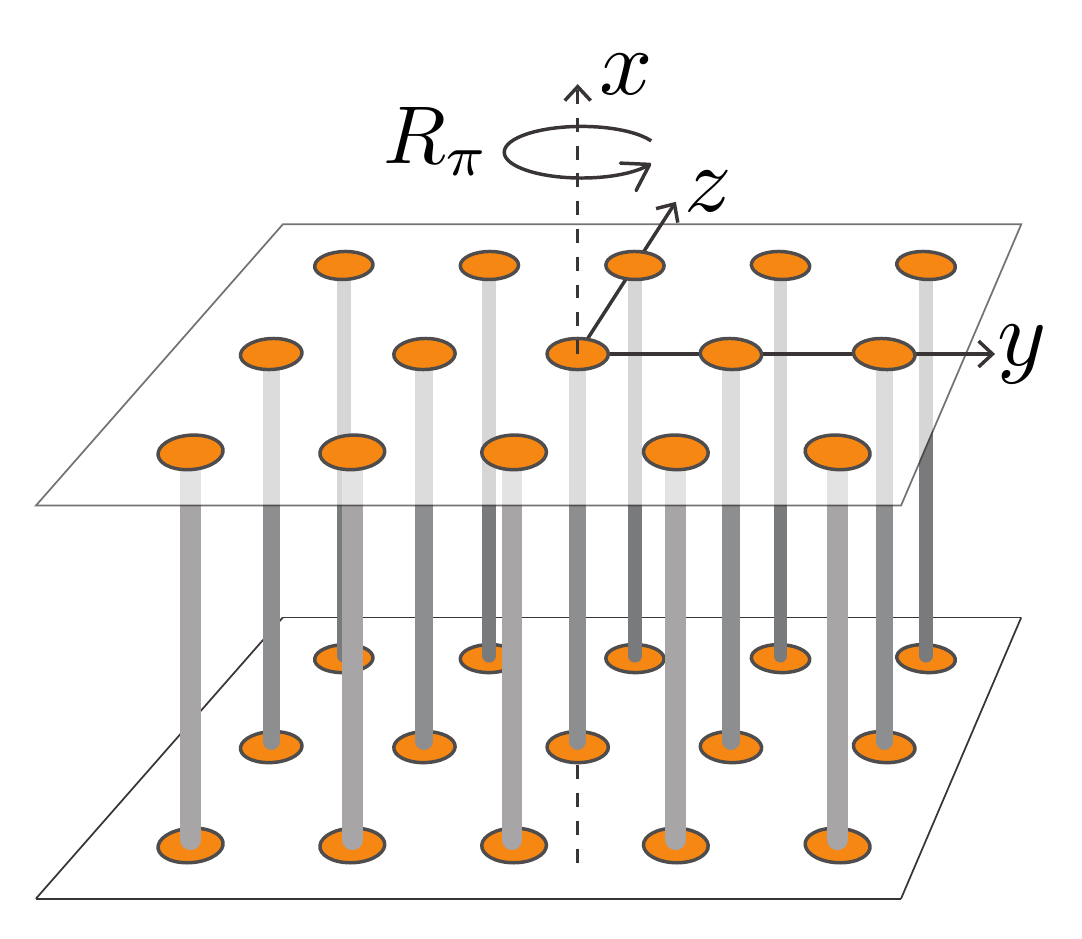}
\caption{The 3D model, which is the 2D array of the 1D chain.
\label{fig4}}
\end{figure}

\section{3D model}
One can readily construct a 3D TCM model in the same way as we did for the 2D models.  The 3D model is a 2D array of the 1D TCM chains, illustrated in Fig.~\ref{fig4}.  For this 3D model, $\mathcal{R}$ must be the site-centered $\pi$-rotation about the $x$-axis. Namely, the rotation axis must coincide with one of the 1D chain.

The gapless 2D surfaces at $x=1$ and $x=L_x$ are protected by the combined symmetry $\mathcal{R}\mathcal{T}$. To see this, let us again add a $\mathcal{R}\mathcal{T}$-symmetric perturbation $\hat{H}'=\sum_{y,z} \vec{H}(y,z)\cdot\hat{\vec{\Gamma}}_{(x=1,y,z)}^1$. To be consistent with the $\mathcal{R}\mathcal{T}$ symmetry, $\vec{H}(y,z)$ should satisfy $\vec{H}(y,z)=-\vec{H}(-y,-z)$, meaning that $H(0,0)=0$. Therefore, there will be a residual zero mode at the ``vortex core'' of the perturbed surface, protected by $(\hat{\mathcal{R}}\hat{\mathcal{T}})^2=-\hat{I}$.

\section{Conclusion}
In this paper we introduced TCM phases protected by non-local symmetry $\mathcal{R}\mathcal{T}$ in two and three dimension. They are interaction-enabled and are robust against attaching physical electrons to the edge.  They can be detected in experiment from their anisotropic response of the edge state towards external magnetic fields.

\begin{acknowledgements}
We thank Yang Qi and Yohei Fuji for insightful discussions. 
LF is supported by the DOE Office of Basic Energy Sciences, Division of Materials Sciences and Engineering under Award No. DE-SC0010526.
\end{acknowledgements}

\end{document}